\documentclass[12pt,preprint]{aastex}
\usepackage{emulateapj5}
\usepackage{apjfonts}
\usepackage{natbib}

\input psfig.tex

\def\aa{{A\&A}}
\def\aas{{ A\&AS}}
\def\aj{{AJ}}
\def\al{$\alpha$}

\def\annrev{{ARA\&A}}
\def\apj{{ApJ}}
\def\apjs{{ApJS}}
\def\asec{$^{\prime\prime}$}

\def\deg{$^{\circ}$}

\def\kms{km s$^{-1}$}
\def\lamb{$\lambda$}
\def\lax{{$\mathrel{\hbox{\rlap{\hbox{\lower4pt\hbox{$\sim$}}}\hbox{$<$}}}$}}
\def\gax{{$\mathrel{\hbox{\rlap{\hbox{\lower4pt\hbox{$\sim$}}}\hbox{$>$}}}$}}
\def\simlt{\lower.5ex\hbox{$\; \buildrel < \over \sim \;$}}
\def\simgt{\lower.5ex\hbox{$\; \buildrel > \over \sim \;$}}

\def\mnras{{MNRAS}}

\def\pasp{{PASP}}

\def\percm2{cm$^{-2}$}

\def\sigmastar{$\sigma_\star$}

\def\heii{\ion{He}{2}}

\def\hii{\ion{H}{2}}
\def\oiii{[\ion{O}{3}]}

\def\oi{[\ion{O}{1}]}
\def\nii{[\ion{N}{2}]}

\def\sii{[\ion{S}{2}]}

\slugcomment{To appear in {\it The Astrophysical Journal Supplement Series}.}

\shorttitle{CATALOG OF STELLAR VELOCITY DISPERSIONS}
\shortauthors{HO ET AL.}

\begin{document}
 
\title{A Search for ``Dwarf'' Seyfert Nuclei. VII. A Catalog of Central Stellar 
Velocity Dispersions of Nearby Galaxies}

\author{Luis C. Ho}

\affil{The Observatories of the Carnegie Institution of Washington, 813 Santa
Barbara St., Pasadena, CA 91101}
 
\author{Jenny E. Greene\altaffilmark{1}}
\affil{Department of Astrophysical Sciences, Princeton University,
Princeton, NJ}
 
\author{Alexei V. Filippenko}
\affil{Department of Astronomy, University of California, Berkeley, CA
94720-3411}

\and

\author{Wallace L. W. Sargent}
\affil{Palomar Observatory, California Institute of Technology, MS 105-24, 
Pasadena, CA 91125}

\altaffiltext{1}{Hubble Fellow, Princeton-Carnegie Fellow.}

\begin{abstract}
We present new central stellar velocity dispersion measurements for 428 
galaxies in the Palomar spectroscopic survey of bright, northern 
galaxies.  Of these, 142 have no previously published measurements, most being 
relatively late-type systems with low velocity dispersions (\lax 100 \kms).  
We provide updates to a number of literature dispersions with large 
uncertainties.  Our measurements are based on a direct pixel-fitting technique 
that can accommodate composite stellar populations by calculating an optimal 
linear combination of input stellar templates.  The original Palomar survey 
data were taken under conditions that are not ideally suited for deriving 
stellar velocity dispersions for galaxies with a wide range of Hubble types.  
We describe an effective strategy to circumvent this complication and 
demonstrate that we can still obtain reliable velocity dispersions for this 
sample of well-studied nearby galaxies.
\end{abstract}


\section{Introduction}

The stellar velocity dispersion (\sigmastar) of the central regions of 
galaxies is a parameter of considerable importance for a variety of 
extragalactic investigations.  Since the early pioneering work of Burbidge et 
al. (1961) and Minkowski (1962), many techniques have been developed for 
measuring \sigmastar\ (e.g., Morton \& Chevalier 1972; Richstone \& Sargent 
1972; Simkin 1974; Sargent et al.  1977; Tonry \& Davis 1979; Bender 1990; 
Rix \& White 1992; van~der~Marel \& Franx 1993; Cappellari \& Emsellem 2004; 
Statler 1995; Barth et al. 2002).  Given the extensive body of observational 
material on \sigmastar\ for nearby galaxies, a number of catalogs have been 
compiled to consolidate the data.  The most widely used of these are the 
catalog of Whitmore et al. (1985), which was updated by McElroy (1995), and 
of Prugniel et al. (1998), which is continuously updated and is available 
through the electronic database HyperLeda (Paturel et al. 2003)\footnote{\tt 
http://leda.univ-lyon1.fr/}.  

The vast majority of the 
published measurements of \sigmastar\ pertain to early-type galaxies, 
largely giant ellipticals and S0s.  Significantly less data are available for 
galaxies along the spiral sequence, and those that have been published often 
show marked disagreement from study to study, as can be seen from perusal of 
the data tabulated in the above-mentioned catalogs.  It is disconcerting that 
many of the highly discrepant entries are, in fact, associated with nearby, 
bright, well-studied galaxies.  The scatter in the published values of 
\sigmastar\ can be blamed, at least in part, on the inherent 
heterogeneity of combining many disparate sources, which often employ 
different telescopes, detectors, apertures, observing strategies, and 
analysis techniques.  The above-cited catalogs attempt to homogenize the final 
compilations by scaling the individual literature sources to a set of 
``standard'' galaxies measured through a roughly constant aperture size 
(2\asec$\times$4\asec).

Notwithstanding these efforts, there is considerable motivation for
assembling an independent, homogeneous, internally consistent set of
new measurements, especially if the data cover a large sample of
galaxies representing a wide range of Hubble types.  A number of previous 
studies have been carried out with this goal in mind, mostly focused on 
relatively early-type galaxies (e.g., Davies et al. 1987; Bernardi et al. 
2003). Our present paper adds to this effort using data taken as part of the 
Palomar spectroscopic survey of nearby galaxies.  During the course of an 
extensive investigation primarily aimed at characterizing the nature of 
nuclear activity in nearby galaxies, we collected high-quality, 
moderate-resolution, long-slit optical spectra of the central regions of 486 
bright, northern galaxies.  The survey was conducted during the period
1984--1990; technical details of the survey and presentation of
various data products and science results can be found in earlier papers in 
this series (Filippenko \& Sargent 1985; Ho et al. 1995, 1997a--1997e, 2003).
This contribution focuses on central stellar velocity dispersions
extracted from the survey.

\section{The Survey}

A full description of the Palomar survey is given by Ho et al. (1995, 1997a).  
Here we mention only a few pertinent details. The survey covers a nearly 
complete, magnitude-limited 

\vskip 0.3cm
\begin{figure*}[t]
\centerline{\psfig{file=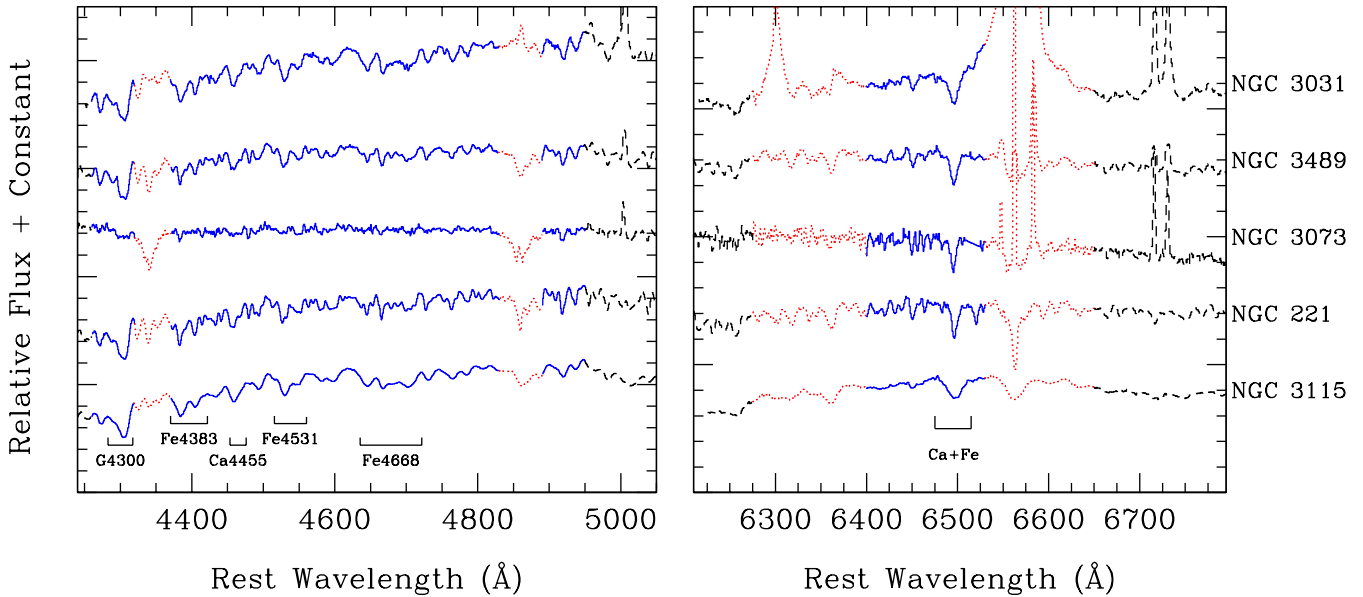,width=19.5cm,angle=270}}
\figcaption[fig1.ps]{
Sample blue ({\it left}) and red ({\it right}) spectra from the Palomar
survey, adapted from Ho et al. (1995).  The intensity of each spectrum has been
scaled and arbitrarily shifted for clarity.  The regions included in the fit
are plotted in blue, while those masked from the fit are plotted as red dotted
lines.  Black dashed lines denote regions outside of the fitting window.
The stellar metal-line indices defined by Ho et al. (1997a) are labeled on the
bottom of each panel.
\label{fig1}}
\end{figure*}
\vskip 0.3cm

\noindent
sample of 486 galaxies from the Revised 
Shapley-Ames catalog (Sandage \& Tammann 1981) that satisfy $B_T \leq 12.5$ 
mag and $\delta > 0$\deg.  The spectra were acquired  using the Double 
Spectrograph (Oke \& Gunn 1982) mounted at the Cassegrain focus of the 
Hale 5-m telescope at Palomar Observatory.  A 2\asec-wide slit was used 
for most of the survey.  The spectra simultaneously cover the regions 
$\sim$4230--5110 \AA\ and $\sim$6210--6860 \AA.  The average full-width at 
half-maximum intensity (FWHM) spectral resolutions on the blue and red sides, 
as determined from comparison-lamp emission lines, are approximately 4.2 \AA\ 
and 2.2 \AA, respectively. These correspond to velocity resolutions, expressed 
as a Gaussian dispersion, of $\sigma_{\rm inst}$ = 118 and 42 \kms\ at 
4500~\AA\ and 6500~\AA, respectively.  (About 10\% of the blue spectra were 
acquired in a slightly higher resolution mode with $\sigma_{\rm inst}$ = 
74 \kms.) The spectra analyzed in this paper are the same as those reported in 
the spectral atlas of Ho et al. (1995); they were extracted from a rectangular 
aperture of size 2\asec$\times$4\asec, which is roughly equivalent to linear 
dimensions of 170 pc $\times$ 350 pc for a median distance of 17.9 Mpc (Ho et 
al. 1997a).

\section{Velocity Dispersions}

\subsection{Method}

Our velocity dispersion measurements are based on the direct pixel-fitting 
method, which, as described by a number of authors (e.g., Rix \& White 1992; 
van~der~Marel 1994; Kelson et al. 2000; Barth et al. 2002), has many of 
advantages compared to more traditional methods based on Fourier or 
cross-correlation techniques.  The Palomar survey has several characteristics 
that pose special challenges for measuring accurate stellar velocity 
dispersions.  First, the majority of the survey galaxies contain emission 
lines from active galactic nuclei (AGNs), often strong and of substantial 
velocity width, presenting a significant source of contamination for the 
stellar absorption features.  Second, the spectral coverage of the survey was 
optimized for obtaining emission-line diagnostics and not for velocity 
dispersion measurements.  Finally, the survey covers a very broad range of 
Hubble types, from dwarf irregulars to giant ellipticals.  Galaxies with a 
wide range of stellar populations are especially susceptible to template 
mismatch.  We use a modified version of the direct pixel-fitting code 
developed by Greene \& Ho (2006).  In brief, a nonlinear Levenberg-Marquardt 
minimization algorithm is used to compare the observed galaxy spectrum with a 
model spectrum $M(\lambda)$, which is assumed to be the convolution of a 
stellar template spectrum, $T(\lambda)$, and a line-of-sight velocity 
broadening function approximated as a Gaussian, $G(\lambda)$:

\begin{equation}
M(\lambda) = P(\lambda)~\{[T(\lambda)~\otimes\ G(\lambda)]+ C(\lambda)\}.
\end{equation}

\noindent
Here, $C(\lambda)$ is an additive term to dilute the stellar features. It can 
be a power-law function to represent an AGN continuum, if present, or any other 
smooth component such as the featureless continuum from hot stars.  For many 
of our later-type galaxies, adding a simple $f_\lambda$ = constant term 
effectively mimics the continuum dilution of the metal lines by 
intermediate-age (A and early-F type) stars in the composite stellar 
population.  The multiplicative factor $P(\lambda)$, typically chosen to be a 
third-order Legendre polynomial, accounts for large-scale mismatches in the 
continuum shapes of the galaxy and template star(s), which can arise from 
internal reddening in the galaxy, stellar population differences, and possible 
residual calibration errors.

An important improvement over the original code of Greene \& Ho is that 
$T(\lambda)$, rather than being a single star, can be an optimal linear 
combination of several stars determined through a nonlinear least-squares fit.
In the case of later-type spirals, especially, this modification provides a 
much better fit for their composite stellar populations, as well as a more 
robust 

\vskip 0.3cm
\begin{figure*}[t]
\centerline{\psfig{file=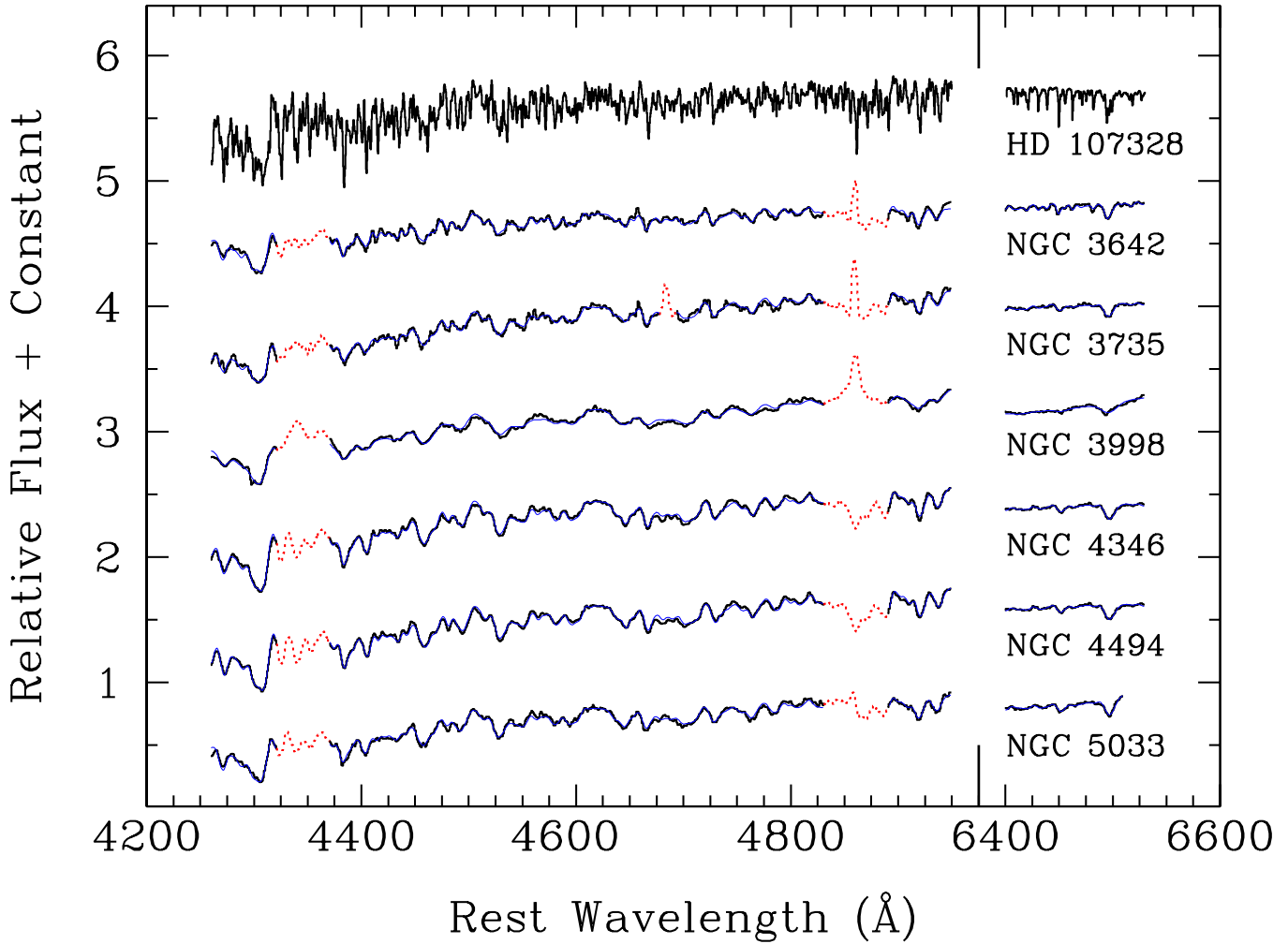,width=19.5cm,angle=270}}
\figcaption[fig2.ps]{
Sample fits for a representative set of galaxies.  The top spectrum is that of
the K0~III star HD~107328 from the Valdes et al. (2004) stellar library.
For each galaxy, the original data are plotted as black histograms.  The
best-fitting model constructed from an optimal combination of broadened
stellar templates is plotted as a thin blue curve.  The regions excluded from
the fit are marked as red dotted lines. The intensity of each spectrum has
been scaled and arbitrarily shifted for clarity.
\label{fig2}}
\end{figure*}
\vskip 0.3cm

\noindent
determination of the final velocity dispersion of the galaxy because the 
intrinsic widths of the template stars vary with spectral type.  Our 
approach of using a mixture of template stars is similar to those employed by 
several previous studies, including Rix \& White (1992) and Cappellari \& 
Emsellem (2004).

\subsection{Fitting Regions}

The blue setup just misses Mg~I \lamb 5175 (``Mg {\it b}''), the feature most 
commonly used to derive velocity dispersions in the visible part of the 
spectrum.  Nevertheless, the blue spectra contain a significant number of 
relatively strong metal-line features, including the G band at 4300 \AA, a 
calcium feature at 4455 \AA, and iron features at 4383, 4531, and 4668 \AA\ 
(Fig.~1, {\it left}; see Table~7 in Ho et al. 1997a for definitions of these 
stellar absorption-line indices).  These metal-line features can be used to 
derive stellar velocity dispersions, so long as they are strong enough in the 
integrated spectrum.  For the blue spectra we fit the region 4260--4950 
\AA; the blue end is chosen to include the G band, while the red end avoids 
the \oiii\ \lamb\lamb 4959, 5007 emission lines.  We mask the regions 
containing H$\gamma$ (4320--4370 \AA) and H$\beta$ (4830--4890 \AA).  In some 
strong emission-line objects, it is necessary to mask a small region around 
\heii\ \lamb 4686.

In practice, the above procedure works well for galaxies with a stellar 
population dominated by stars of spectral type mid-F and later, but not for 
those with younger populations.  As Figure~1 illustrates, the spectrum of 
NGC~3073 contains mostly light from stars of type A and early-F, and the 
metal-line features, although clearly present in this spectrum of fairly high 
signal-to-noise ratio (S/N), are significantly diluted by the blue 
continuum of the hotter stars.  Spectra like that of NGC~3073 (which, 
curiously, is an S0 galaxy) typically characterize many of the later-type 
spirals in the survey.  The moderate resolution of the blue spectra presents 
another severe limitation.  Even for galaxies where the blue metal-line 
features are strong and unambiguously detected (e.g., NGC~221 and NGC~3489 in 
Fig.~1), the derived velocity dispersions may be subject to large systematic 
uncertainties if the true dispersions are near or below the native spectral 
resolution of $\sigma_{\rm inst} \approx 120$ \kms.  For example, 

\vskip 0.3cm
\begin{figure*}[t]
\centerline{\psfig{file=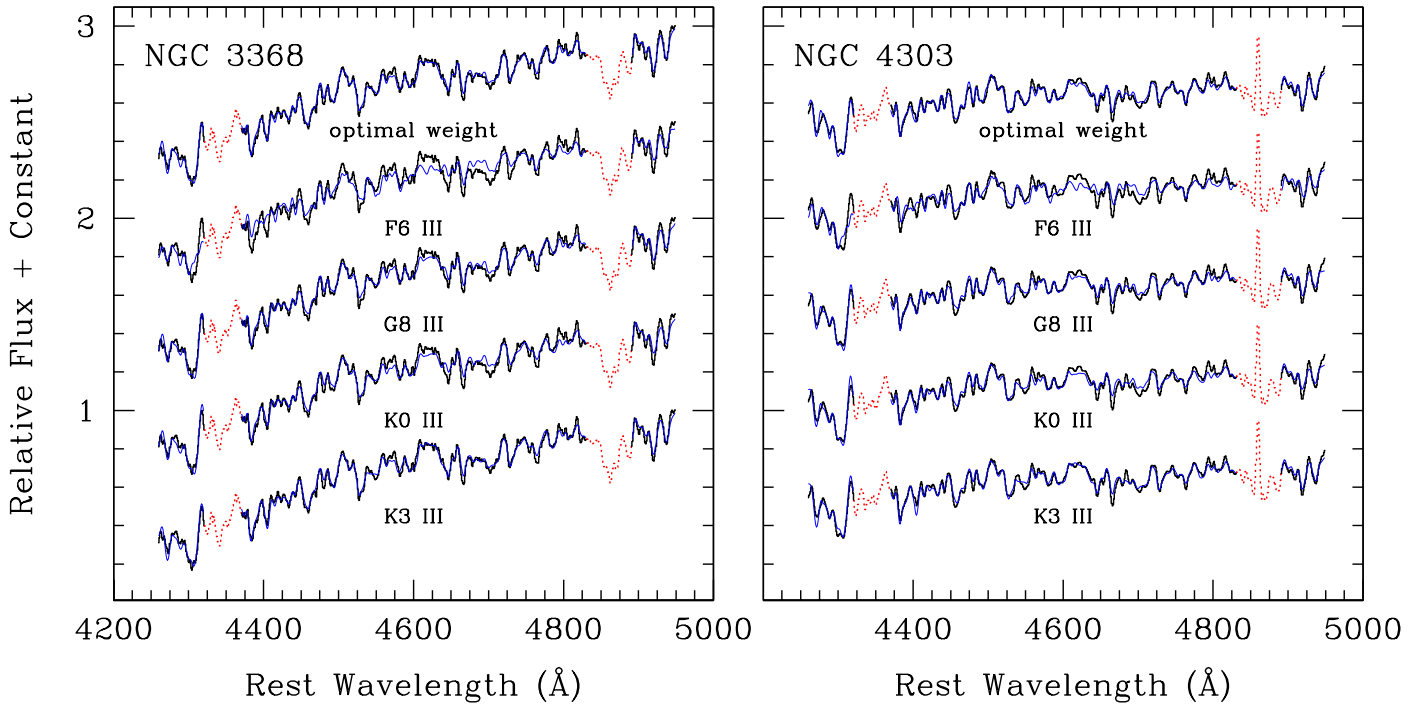,width=19.5cm,angle=270}}
\figcaption[fig3.ps]{Sample fits for NGC~3368 and NGC~4303 in the blue
spectral region.  The top spectrum shows the optimally weighted fit, followed
by fits using single stars of spectral type F6~III, G8~III, K0~III, and
K3~III.  The original data are plotted as black histograms, the fits are
plotted as blue curves, and the regions excluded from the fit are plotted as
red dotted lines.  The intensity of each spectrum has been scaled and
arbitrarily shifted for clarity.
\label{fig3}}
\end{figure*}
\vskip 0.3cm

\noindent
according to 
the literature NGC~221 and NGC~3489 have $\sigma = 72$ and 112 \kms, 
respectively.

The red spectra, with $\sigma_{\rm inst} \approx 40$ \kms, provide crucial 
relief to the many galaxies in the survey that suffer from insufficient 
resolution in the blue.  Unfortunately, very few strong, uncontaminated 
stellar features exist in the spectral coverage of our red setup, which is 
dominated almost entirely by strong emission lines (\oi\ \lamb\lamb 6300, 6363, 
\nii\ \lamb\lamb 6548, 6583, H\al, and \sii\ \lamb\lamb6716, 6731).  One 
glimmer of hope lies with the Ca+Fe feature at 6495 \AA.  To the best of our 
knowledge, this little-known feature has never been used explicitly for 
kinematical measurements in galaxies, although it has played a role in
other contexts such as the determination of radial-velocity curves for
the secondary stars in black hole X-ray binaries (e.g., Filippenko et al.
1995, 1997).  We will show that it plays a central role in our survey.  

As Figure~1 ({\it right}) illustrates, the Ca+Fe feature, lying just blueward 
of the H\al+\nii\ complex, is fairly well isolated, even in objects with 
prominent, broad H\al\ emission (e.g., NGC~3031).  Importantly, it is 
moderately strong in nearly all galaxies, even those whose blue spectra are 
hopeless diluted by A and F-type stars (e.g., NGC~3073).  Using the 
measurements published by Ho et al. (1997a, Table~9), we find that Ca+Fe was 
reliably detected in 438 out of the 486 galaxies in the Palomar survey (90\%), 
with an average equivalent width of $\langle{\rm W(Ca+Fe)}\rangle = 0.9$ 
\AA.  There is, at most, a factor of 2 variation in line strength from one 
extreme end of the Hubble sequence to the other.  Among ellipticals and S0s 
(morphological type index $-6 < T < 0$; de~Vaucouleurs et al. 1991), 
$\langle{\rm W(Ca+Fe)}\rangle = 1.2$ \AA, to be compared with 
$\langle{\rm W(Ca+Fe)}\rangle = 0.6$ \AA\ for Sc--Sdm spirals (morphological 
type index $5 < T < 9$).

After some experimentation, we find that the most stable fitting region for the
red setup is 6400--6530 \AA\ (Fig.~1, {\it right}).  The blue limit provides 
as much leverage as possible to define the continuum level without colliding 
with \oi\ \lamb 6363, and the red limit abuts \nii\ \lamb 6548.  In a few 
objects with very strong, broad H\al\ emission, we had to curtail the red 
limit to 6510 \AA; in these cases, it was often also helpful to increase the 
order of the polynomial factor (to $\sim 5-6$) to better trace the 
steeply rising gradient of the blue wing of the H\al\ emission line.

\subsection{Template Stars}

In addition to spectrophotometric standard stars, during the course of the 
survey we usually also took nightly observations of at least one late-type 
giant star to be used as a velocity template.  Velocity standards were 
not observed in a small number of observing runs; this affected 50 galaxies, 
or roughly 10\% of the survey.  Because measuring velocity dispersions was not 
a top priority for the original survey, neither the number of stars nor their 
range of spectral types was chosen optimally. In some of the runs, only a 
single velocity template was observed, and at most there were two.

The limitations of the Palomar template stars compel us to explore an 
alternative calibration strategy.  We use as our primary source of templates 
the library of Coud\'e-feed stellar spectra published by Valdes et 
al. (2004).  This tremendously useful database contains high-S/N spectra of 
1273 stars of essentially all spectral types, covering 3460 to 9464 \AA.  The 
spectral resolution of the library, FWHM $\approx$ 1 \AA, is significantly 
higher than that of either the blue or red Palomar spectra.  Thus, the Valdes 
stars can be used as velocity templates for the Palomar galaxies, after
accounting for the differential instrumental broadening 

\vskip 0.3cm
\begin{figure*}[t]
\centerline{\psfig{file=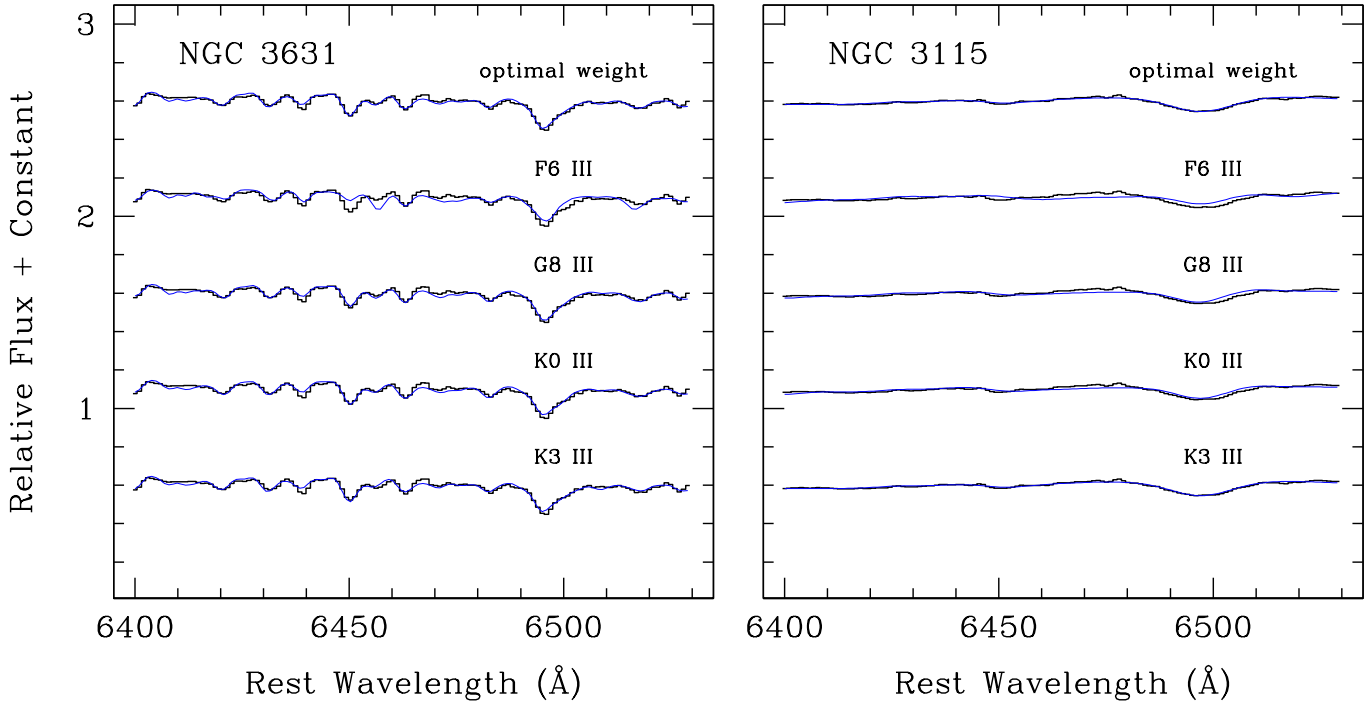,width=19.5cm,angle=270}}
\figcaption[fig4.ps]{Sample fits for NGC~3631 and NGC~3115 in the red
spectral region.  The top spectrum shows the optimally weighted fit,
followed by fits using single stars of spectral type F6~III, G8~III, K0~III,
and K3~III.  The original data are plotted as black histograms, and the fits
are plotted as blue curves.  The intensity of each spectrum has been scaled and
arbitrarily shifted for clarity.
\label{fig4}}
\end{figure*}
\vskip 1.3cm

\noindent
between the two 
data sets.   

The Valdes library also gives us an extensive selection of stars 
of different spectral types for our optimal fit.  Through experimentation, we 
find that in general a set of four stars---spectral types F6~III, G8~III, 
K0~III, and K3~III---suffices to account for the stellar population mixture of 
almost all galaxies in our sample.  We give preference to stars of near-solar 
metallicity to try to approximate the conditions in galactic bulges.  Although 
type-A and early-F stars clearly exist in some galaxies, in practice they do 
not need to be included because our fitting regions deliberately avoid the 
Balmer absorption lines (Fig.~1) and the continuum dilution term [$C(\lambda)$ 
in Equation 1] effectively mimics the hot continuum of these stars.

\subsection{Fitting Results}

Figure~2 gives examples of some typical fits.  The top spectrum is that of 
the red giant (K0~III) star HD~107328, shown to help guide the eye to identify 
the stellar features.  Subsequent spectra illustrate galaxies with a wide 
range in emission-line strengths and velocity dispersions.  The original 
galaxy spectrum is plotted as black histograms; the best-fitting, optimally 
weighted, broadened velocity template is plotted as a thin blue line; and the 
masked regions are plotted as a red dotted line.  Using a set of just four 
stars, we can usually achieve quite good fits, with formal statistical 
errors on the velocity dispersions in the range of 5\%--10\%.  The results are 
also quite robust with respect to the choice of template stars; interchanging 
different stars of the same spectral type and similar metallicity affects the 
final dispersions at the level of $1\%$ or less.  In most objects, the largest 
fraction of the light comes, not surprisingly, from K giants. The Fe 
$\lambda$4668 feature, in particular, is very sensitive to K1~III--K3~III 
giant stars, which significantly improve the fit over the region 4600--4800 
\AA\ (Fig.~3).  Our fitting region for the red setup, especially the Ca+Fe 
\lamb6495 feature, is also very sensitive to K1~III--K3~III giants (Fig.~4).  
The vast majority of the galaxies, however, including many bulge-dominated 
systems, require some contribution from G and even F-type stars.

To translate the Valdes-based dispersions onto the Palomar system, we subtract 
in quadrature the relative resolution difference between the Valdes and 
Palomar systems.  Assuming the nominal instrumental resolutions of the two 
data sets, the resolution correction for the blue side is $\sigma_c = 114.8 
\pm 5.8$ \kms\ ($68.4 \pm 7.1$ \kms\ for the higher resolution mode), while 
that for the red side is $\sigma_c = 37.4 \pm 7.5$ \kms, where the error bar 
represents the root-mean square (rms) scatter of the night-to-night 
variations of the Palomar instrumental resolution.  The validity of this simple 
approach can be verified empirically by comparing the corrected dispersions 
with published values.  Among the 223 galaxies with velocity dispersions 
derived in the blue, 189 have literature measurements; of the 422 dispersions 
measured in the red, 283 have literature values.  As illustrated in Figure~5, 
the adopted resolution corrections yield reasonably satisfactory agreement 
between our dispersion measurements and the literature values, particularly in 
the regime when the dispersions are well resolved ($\sigma$ \gax\ 
$\sigma_{\rm inst}$; {\it solid points}).  On the blue side (Fig.~5{\it a}), 
for $\sigma$ \gax\ $\sigma_{\rm inst} \approx 120$ \kms, 
$\langle\sigma_{\rm blue}-\sigma_{\rm Literature}\rangle = 1.2$ \kms\ with an 
rms scatter of 25.3 \kms.  The red side delivers useful measurements down to 
$\sigma \approx \onehalf \sigma_{\rm inst} \approx 20$ \kms\ (Fig.~5{\it b}).  
Over the entire velocity range, 
$\langle\sigma_{\rm red}-\sigma_{\rm Literature}\rangle = 3.0$ \kms\ with an 
rms scatter of 28.3 \kms.  There is no perceptible systematic bias, provided 
that the optimal fit excludes the K3~III star, as explained below.

Our initial fits for the red-side spectra, which include the full 

\vskip 0.3cm
\begin{figure*}[t]
\centerline{\psfig{file=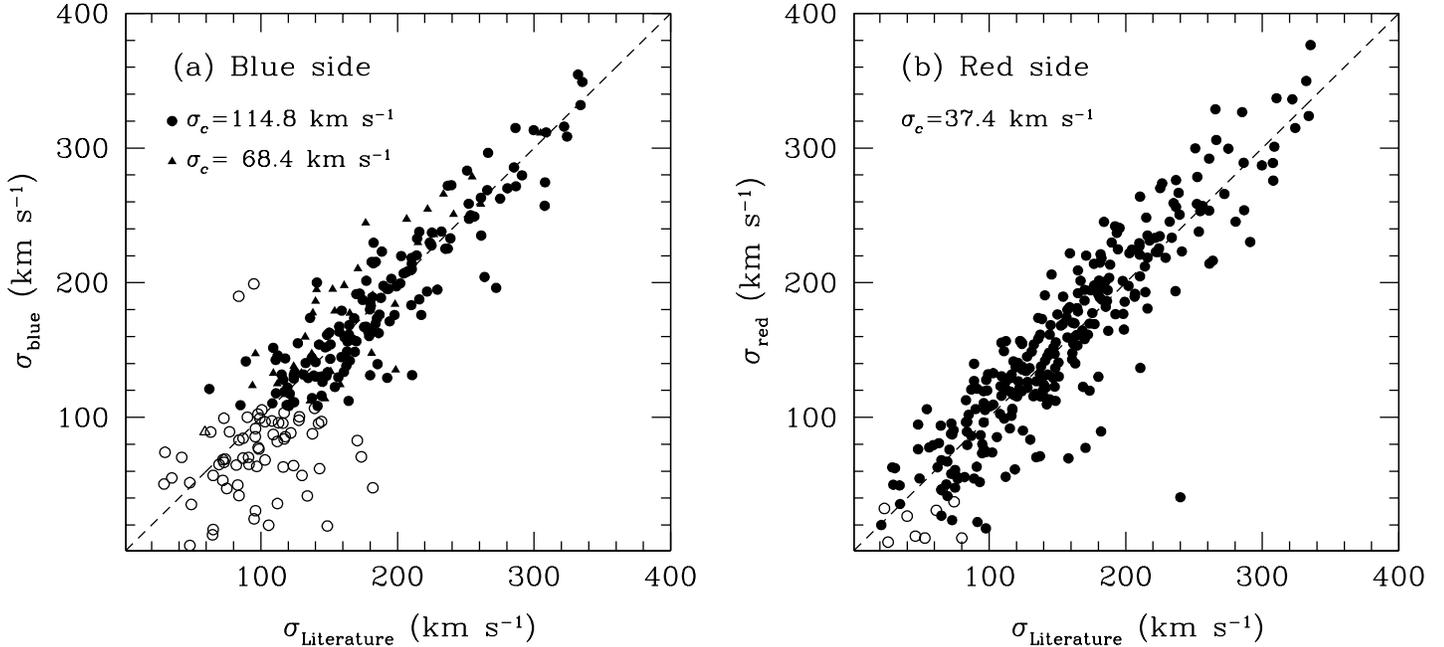,width=19.5cm,angle=270}}
\figcaption[fig5.ps]{
Comparison between velocity dispersions published in the literature with
velocity dispersions derived using an optimal combination of Valdes template
stars for the ({\it a}) blue side and ({\it b}) red side, corrected for the
relative resolution difference ($\sigma_c$) between the Valdes and Palomar
systems (see \S 3.3).  Open symbols mark objects that are poorly resolved.
The dashed diagonal line denotes equality.
\label{fig5}}
\end{figure*}
\vskip 1.3cm

\noindent
complement of 
four template stars (F6~III to K3~III), revealed a puzzling systematic trend.  
Whereas the fits for low-$\sigma$ galaxies yield dispersions that, after 
resolution correction, agree reasonably well with literature values, objects 
with $\sigma$ \gax\ 150--200 \kms\ show a net systematic offset toward larger 
velocities, by roughly $+30$ \kms.  We believe that this effect arises from
template mismatch.  As shown in Figure~4, in small, low-luminosity bulges, 
such as that in the Sc galaxy NGC~3631, the red absorption features, especially Ca+Fe, are nearly equally well fit by template stars of spectral type G8~III, 
K0~III, or K3~III.  In stark contrast, NGC~3115, a luminous S0 galaxy with a 
substantial bulge, clearly singles out the K3~III star as the preferred 
template, which then contributes most of the weight to the optimal fit.  
(We have verified that K1~III and K2~III templates give almost equally good 
fits as the K3~III template.)  Why?  This is because the Ca+Fe feature is 
strongest in high-$\sigma$ galaxies and in late-type giants.  Within the 
Palomar galaxy sample, the strength of the Ca+Fe feature increases roughly 
with velocity dispersion, albeit with significant scatter.  Dividing the 
sample into two, galaxies with $\sigma < 150$ \kms\ have 
$\langle{\rm W(Ca+Fe)}\rangle = 0.87$ \AA, to be compared with 
$\langle{\rm W(Ca+Fe)}\rangle = 1.23$ \AA\ for galaxies with $\sigma \ge 150$ 
\kms.  At the same time, the strength of the Ca+Fe feature in stars increases 
toward later spectral types.  To demonstrate this, we measured the Ca+Fe 
feature for individual stars in the Valdes library, using the index definition 
given in Ho et al. (1997a).  Choosing 15 stars of roughly similar 
metallicities for each spectral type, we find $\langle{\rm W(Ca+Fe)}\rangle = 
0.79$, 0.99, 1.17, 1.35, and 1.53 \AA\ for G8~III, K0~III, K1~III, K2~III, and 
K3~III, respectively.   Galaxies with $\sigma \ge 150$ \kms\ have Ca+Fe 
strengths very similar to those of K1~III--K3~III stars, and thus it is not 
surprising that an optimal fit would give these stars greatest weight.  A bias 
in the derived velocity dispersion for high-$\sigma$ galaxies arises {\it if}\ 
in these systems their Ca+Fe feature is boosted because of an abundance 
enhancement.  We speculate that the 
culprit is Ca.  As an $\alpha$ element, Ca may be enhanced similarly as Mg in 
early-type galaxies (Prochaska et al. 2005; but see Graves et al. 2007).  In 
such a situation, the apparently good match with the K1~III--K3~III templates 
is only an artifact of their mutually strong Ca+Fe feature.  Since such 
late-type giants have very narrow intrinsic line widths, the inferred velocity 
dispersion would be overestimated, thus leading to the observed bias.  To 
bypass this complication, we removed the K3~III giant from the optimal fit of 
the red-side spectra.  

For each galaxy, we compute a final velocity dispersion as the average of the 
blue-side and red-side dispersions, weighted by their respective error bars.
The error bars reflect the quadrature sum of the formal statistical uncertainty
from the optimal fit and the rms scatter of the resolution correction, which
is dominated by the uncertainty in the original instrumental resolution of the 
Palomar spectra.  Among the 428 galaxies with new velocity dispersion 
measurements, 286 have published literature values.  Comparison between the 
objects in common (Fig.~6) show very good consistency.  Over the entire range 
in velocities, $\langle\sigma_{\rm final}-\sigma_{\rm Literature}\rangle=3.0$ 
\kms.  The scatter is still quite large (rms 28.3 \kms), but its magnitude is 
consistent with that found by Barth et al. (2002) based on a smaller sample of 
$\sim 30$ galaxies with high-quality velocity dispersion measurements.  

There are several notable outliers in Figure~6, for which the literature 
values are larger than ours by more than $\sim 80$ \kms.  The most extreme 
case is NGC~520, for which HyperLeda reports $\sigma = 240\pm25$ \kms\ whereas 
we determine $\sigma=40.6\pm8.9$ \kms.  This is a complex, interacting galaxy 
(Arp~157), and the HyperLeda value of $\sigma = 240$ \kms\ pertains to the 
``southeast-northwest'' component, not the primary nucleus of the ``east-west''
component (using the naming convention of Stanford \& 

\vskip 0.3cm
\psfig{file=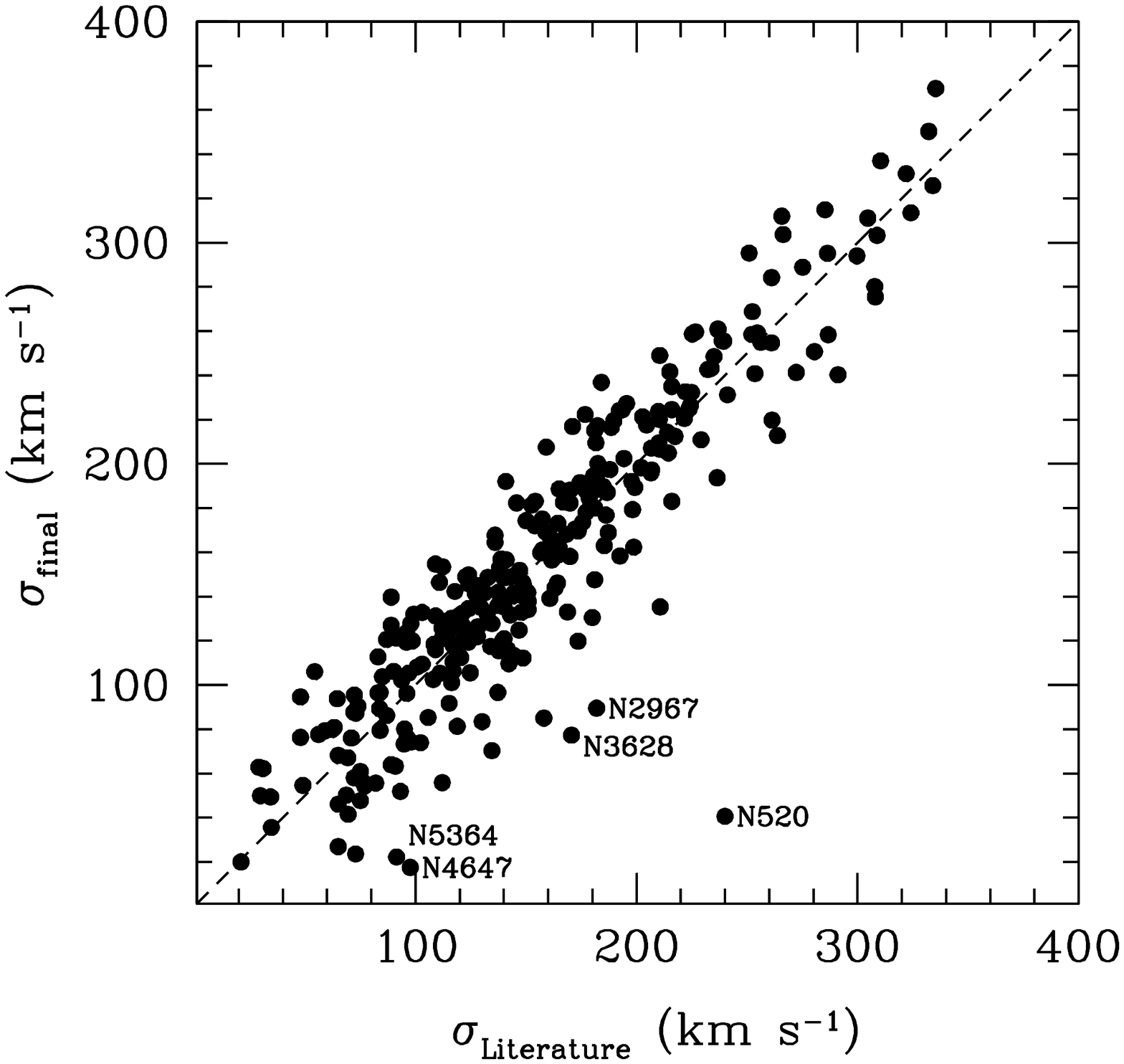,width=8.5cm,angle=0}
\figcaption[fig6.ps]{
Comparison of final velocity dispersions with literature values.
The dashed diagonal line denotes equality.  Several prominent outliers
are labeled (see \S 3.4).
\label{fig6}}
\vskip 0.3cm

\noindent
Balcells 1990a).  The 
Palomar spectrum was centered on the position of the primary nucleus.  From 
visual inspection of the plots published by Stanford \& Balcells (1990a, 
1990b), it appears that the published dispersion of the primary nucleus 
should be $\sigma \approx 100\pm25$ \kms.  (We thank the referee for making 
this estimate for us, with which we agree.) The velocity dispersion for 
NGC~2967 (prior to homogenization), $200\pm27$ \kms, seems suspiciously high 
for an Sc galaxy; according to HyperLeda, it derives from an unpublished 
measurement by B. C. Whitmore \& E. Malumuth (1984).  The same applies to 
the Sc galaxy NGC~4647, for which Hyperleda lists $\sigma = 98\pm39$ \kms.  
Finally, we note that the literature values of both NGC~3628 ($\sigma = 
171\pm71$ \kms) and NGC~5364 ($\sigma = 91\pm52$ \kms) have exceptionally 
large error bars.  If we exclude these five outliers from our sample, 
$\langle\sigma_{\rm final}-\sigma_{\rm Literature}\rangle = 2.7$ \kms, 
and the scatter reduces to 23.6 \kms.

\section{The Catalog}

The final results are presented in Table~1.  For each galaxy, we list the 
literature value of the central stellar velocity dispersion, if available, 
followed by the dispersions derived from the blue ($\sigma_{\rm blue}$) and 
red ($\sigma_{\rm red}$) Palomar spectra, the final value 
($\sigma_{\rm final}$) obtained from the weighted average of 
$\sigma_{\rm blue}$ and $\sigma_{\rm red}$, 
and lastly the adopted value.  Most of the literature values come from the 
HyperLeda database (Paturel et al. 2003), which, for any given galaxy, 
attempts to homogenize all published measurements into a single value by 
applying scaling factors determined from a set of ``standard'' galaxies 
measured through a roughly constant aperture size of 2\asec$\times$4\asec.  
This aperture size, fortunately, exactly matches that employed in the Palomar 
survey.  

For the final, adopted dispersion, there are strong reasons to prefer the 
Palomar measurements because of their homogeneity.  Although in many cases
their error bars formally exceed those of the literature sources, we believe 
that the error budget for the Palomar measurements is realistic, as evidenced, 
for example, from comparison with the high-accuracy measurements from Barth et 
al. (2002) for galaxies in common.  Nevertheless, for concreteness, the final 
column of Table~1 lists either the final Palomar dispersion or the literature 
value, if available, based on whichever has the smaller formal error bar.

In total, our catalog gives new stellar velocity dispersion measurements 
for 428 galaxies, 88\% of the parent survey.  Of these, 142 (30\%) have no 
previously published measurements.  Not surprisingly, most of the new 
measurements are for late-type galaxies, systems where velocity dispersions
are more challenging to obtain because of their characteristically lower 
values (\lax 100 \kms) and complications due to their composite stellar 
populations and contamination by emission lines.  Our new measurements also
provide updates to a number of literature dispersions that previously had 
large uncertainties or, in some instances, were grossly in error.

Stellar velocity dispersions could not be derived for 58 galaxies, mostly 
because their stellar features are too weak.  For the sake of completeness, 
for the 34 of these objects that have emission lines, and for which no 
reliable dispersions exist in the literature, we list an indirect 
estimate of their stellar velocity dispersion based on their observed 
{\it gaseous}\ velocity dispersion derived from the line profile of \nii\ 
\lamb 6583.  Using the current database, Ho (2009) finds that the kinematics 
of the ionized gas in the central few hundred parsecs of bulges generally 
trace the kinematics of the stars, such that $\sigma_g \approx (0.8-1.2) 
\sigma_*$.  In detail, the normalization of the $\sigma_g-\sigma_*$ 
relation shows a slight dependence on nuclear (H\al) luminosity and Eddington
ratio, but {\it only}\ for sources spectroscopically classified as AGNs 
(LINERs, transition objects, and Seyferts).  Those classified as \hii\ 
(star-forming) nuclei obey $\sigma_g = 0.83 \sigma_*$ with an rms 
scatter 0.19 dex.  This is the relation that we use because all of the 34 
emission-line sources with very weak stellar features are \hii\ nuclei (Ho et 
al. 1997a)\footnote{In detail, Ho (2009) notes that $\sigma_g/\sigma_*$ for 
\hii\ nuclei depends on $\sigma_*$, but for our present purposes we neglect 
this complication.}.  The error bars in the adopted dispersions come from the 
quadrature sum of the uncertainties in the original \nii\ line widths (we 
conservatively assume 10\%; Ho et al. 1997a) and the 0.19 dex scatter in the 
$\sigma_g-\sigma_*$ relation.

\section{Summary}

The Palomar spectroscopic survey has furnished considerable insights into the 
nature of nuclear activity in nearby galaxies (see Ho 2008 for a review).  
Aside from some considerations of the central stellar populations (Ho et al. 
2003; Zhang et al. 2008), however, comparatively little analysis has been 
done on the absorption-line component of the spectra.  This paper utilizes the 
survey spectra to derive a homogeneous set of new central stellar velocity 
dispersion measurements.  A major obstacle is that the original survey data 
were not taken with this application in mind.  In particular, neither the 
number nor the range of calibration template stars is ideally suited for 
deriving stellar velocity dispersions for galaxies with a wide range of 
composite stellar populations.  The wavelength coverage of the blue-side and 
red-side spectra is nonstandard for velocity dispersion work and is rather 
sensitive to template mismatch.  Moreover, the spectral resolution of the 
blue-side spectra is too coarse to yield reliable dispersions for most of the 
later-type galaxies in the sample.

We describe an effective strategy to address these challenges.  We use the 
extensive Coud\'e-feed spectral library of Valdes et al. (2004) as the primary
source of stellar templates.  Applying a simple correction for the nominal
relative resolution difference between the Valdes and Palomar systems yields
velocity dispersions that show reasonably good agreement with literature 
values.  The direct-pixel fitting code of Greene \& Ho (2006) was adapted to 
solve for an optimally weighted linear combination of template stars, a 
crucial step to match the composite stellar population typically found in 
later-type galaxies.  We demonstrate that the Ca+Fe \lamb6495 feature in the 
red-side spectra can be used to derive robust velocity dispersions, a crucial 
consideration because the resolution of the red setup, significantly higher 
than that of the blue setup, is sufficient to probe even the late-type systems 
in the survey.

Our final catalog lists a uniform set of new stellar velocity dispersions for
428 galaxies in the Palomar survey.  A significant fraction of the galaxies,
especially later-type systems, have no previously published velocity
dispersions.  Together with indirect estimates for another 34 objects and 
supplementary data from the literature, essentially all (482/486) of the 
galaxies in the Palomar survey now have central velocity dispersion 
measurements.  The Palomar galaxies have been and continue to be heavily
investigated for a variety of scientific applications.  The catalog of velocity
dispersions presented in this paper will add an important new dimension
to the already rich database available for this much-studied galaxy sample.

\acknowledgments

We are very grateful to the staff of Palomar Observatory for their assistance 
with the observations over many years.  We thank Aaron Barth and Dan Kelson 
for advice on methods of measuring velocity dispersions.  An anonymous referee 
offered many helpful and critical comments that improved the paper.  We 
acknowledge the usage of the HyperLeda database 
({\tt http://leda.univ-lyon1.fr}).  The research of L.C.H. was supported by
the Carnegie Institution of Washington.  Support for J.E.G. was provided by
NASA through Hubble Fellowship grant HF--01196, awarded by the Space Telescope
Science Institute, which is operated by the Association of Universities for
Research in Astronomy, Inc., for NASA, under contract NAS 5--26555. A.V.F. 
and W.L.W.S. are grateful for the financial support of the NSF, through grants
AST--0607485 and AST--0606868, respectively.

\clearpage
\begin{figure*}[t]
\centerline{\psfig{file=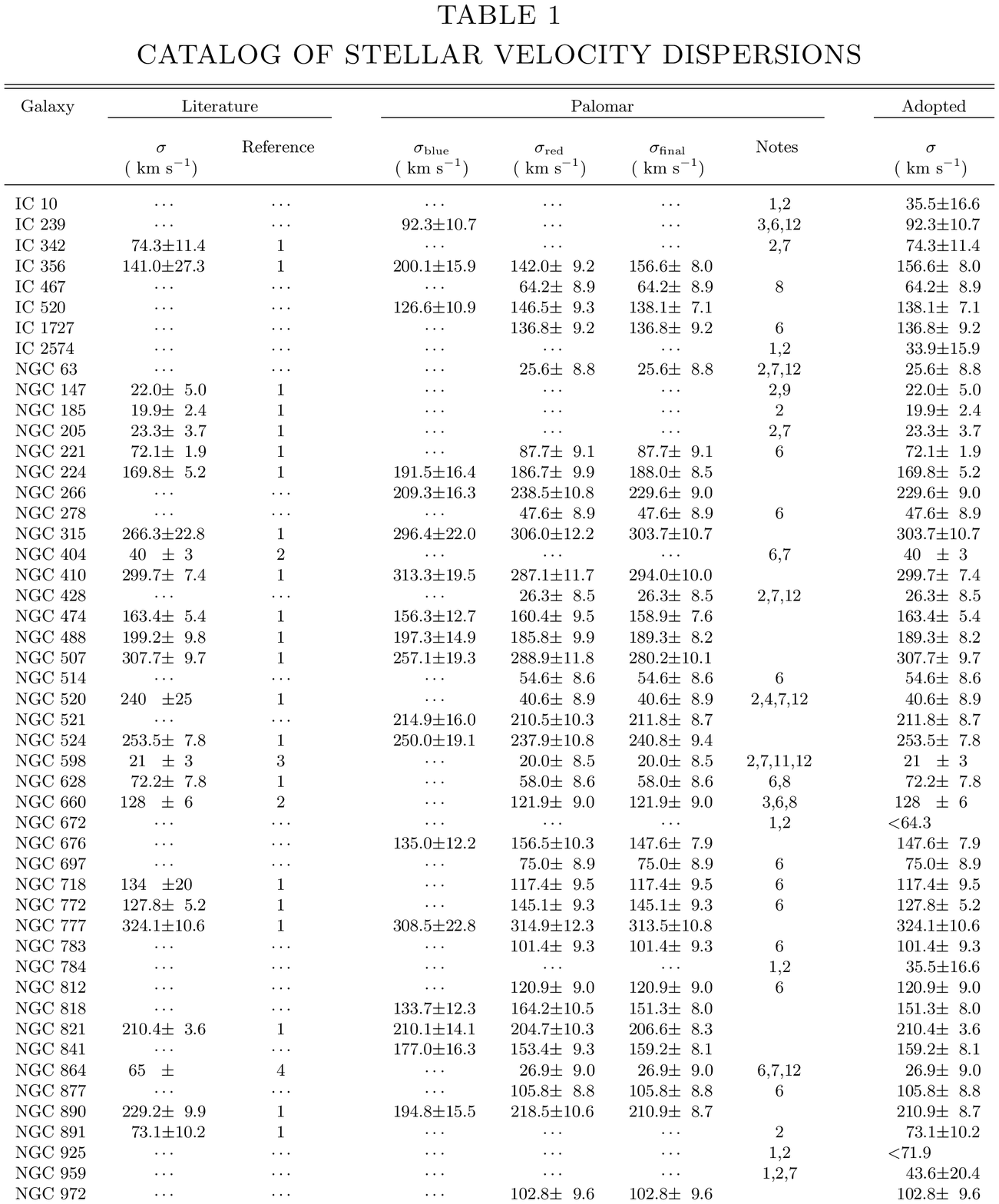,width=18.5cm,angle=0}}
\end{figure*}

\clearpage
\begin{figure*}[t]
\centerline{\psfig{file=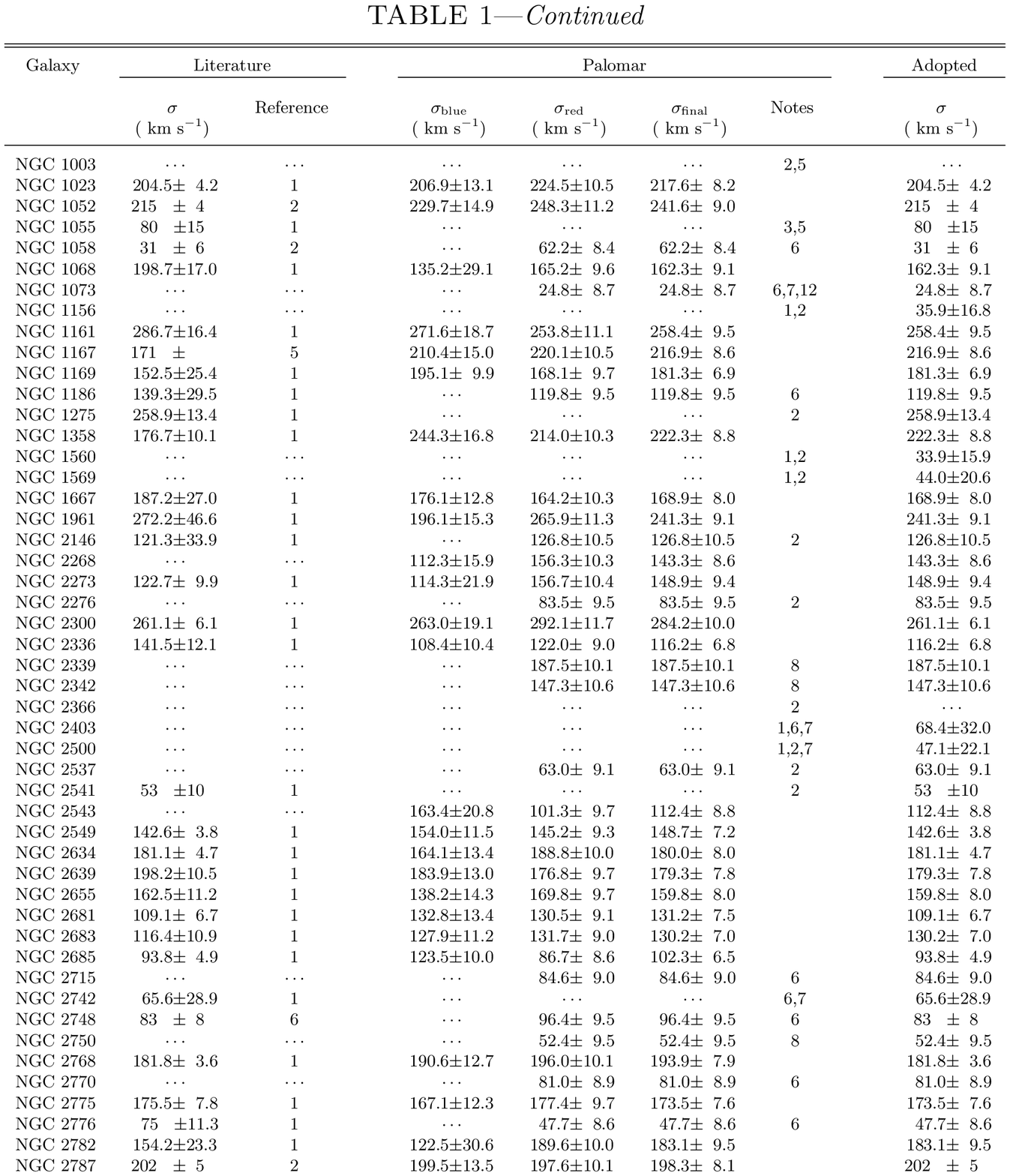,width=18.5cm,angle=0}}
\end{figure*}

\clearpage
\begin{figure*}[t]
\centerline{\psfig{file=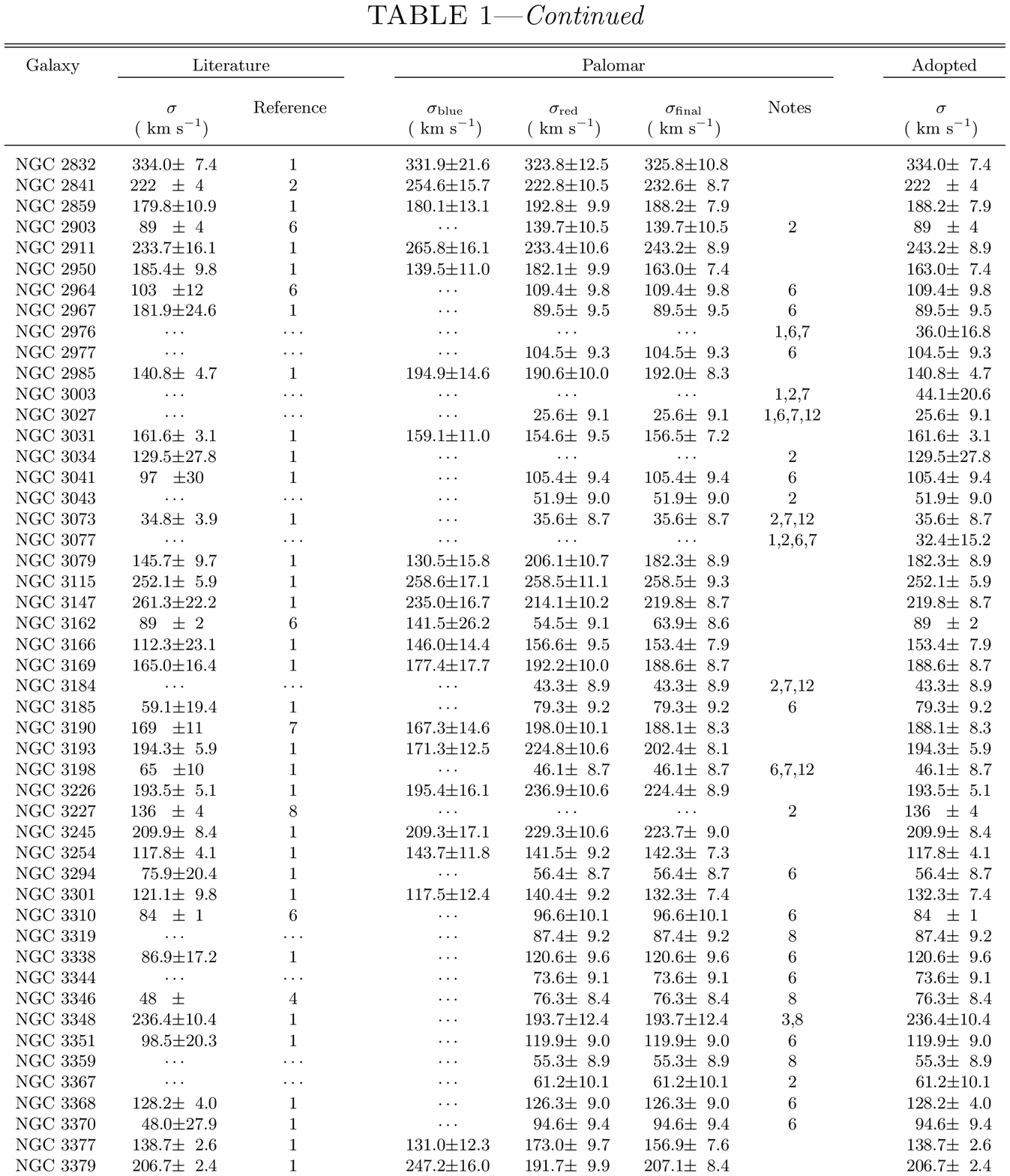,width=18.5cm,angle=0}}
\end{figure*}

\clearpage
\begin{figure*}[t]
\centerline{\psfig{file=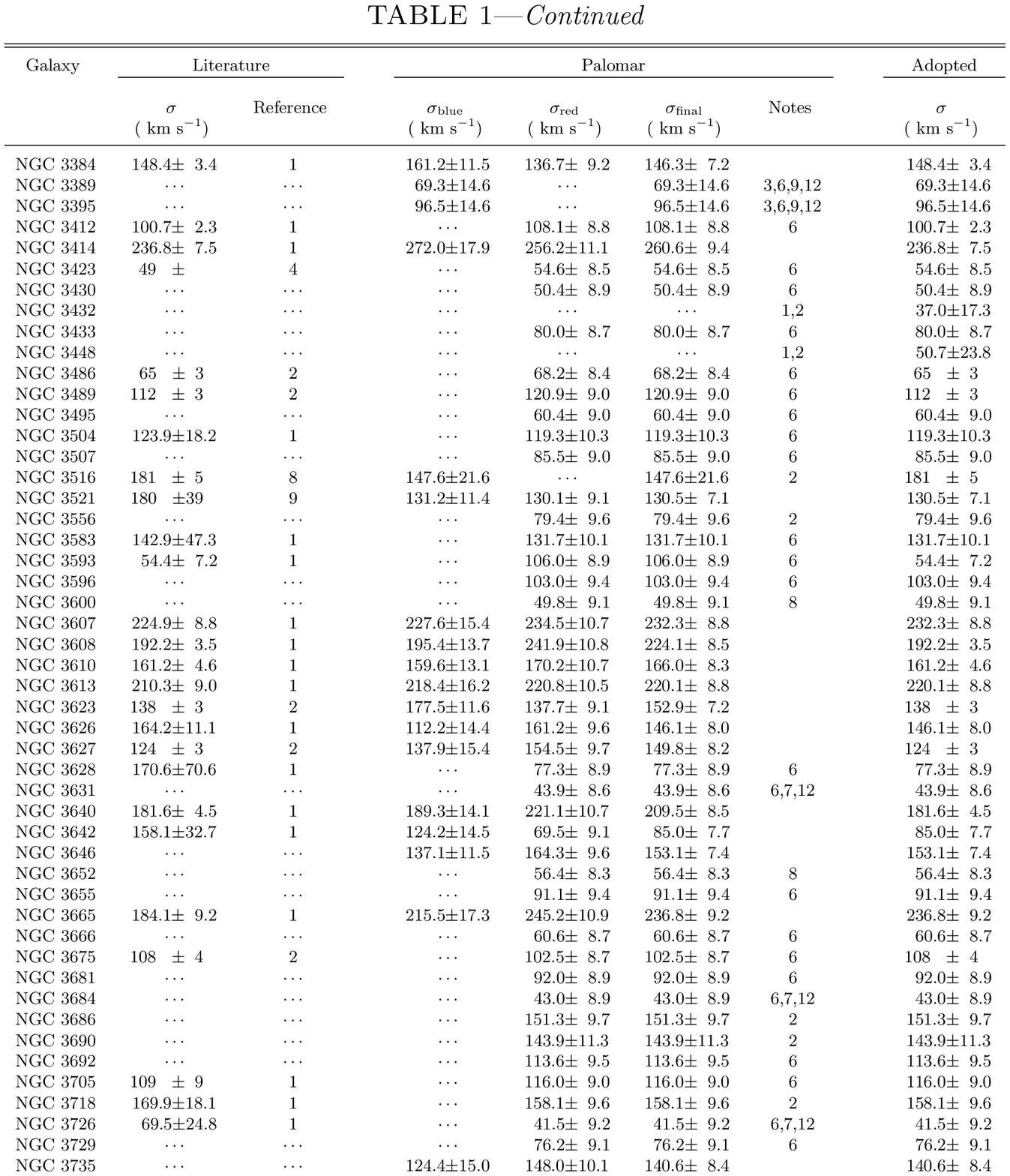,width=18.5cm,angle=0}}
\end{figure*}

\clearpage
\begin{figure*}[t]
\centerline{\psfig{file=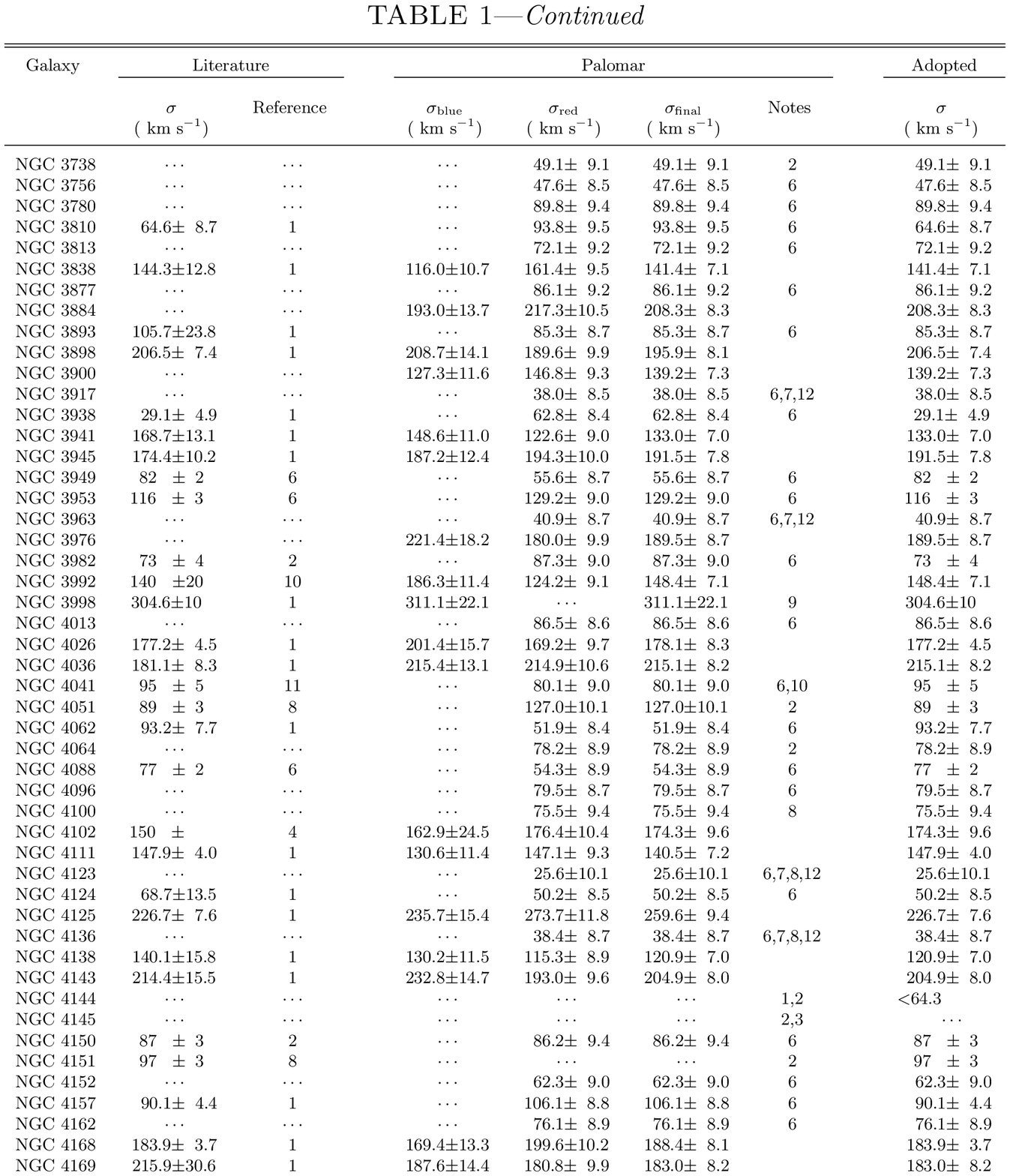,width=18.5cm,angle=0}}
\end{figure*}

\clearpage
\begin{figure*}[t]
\centerline{\psfig{file=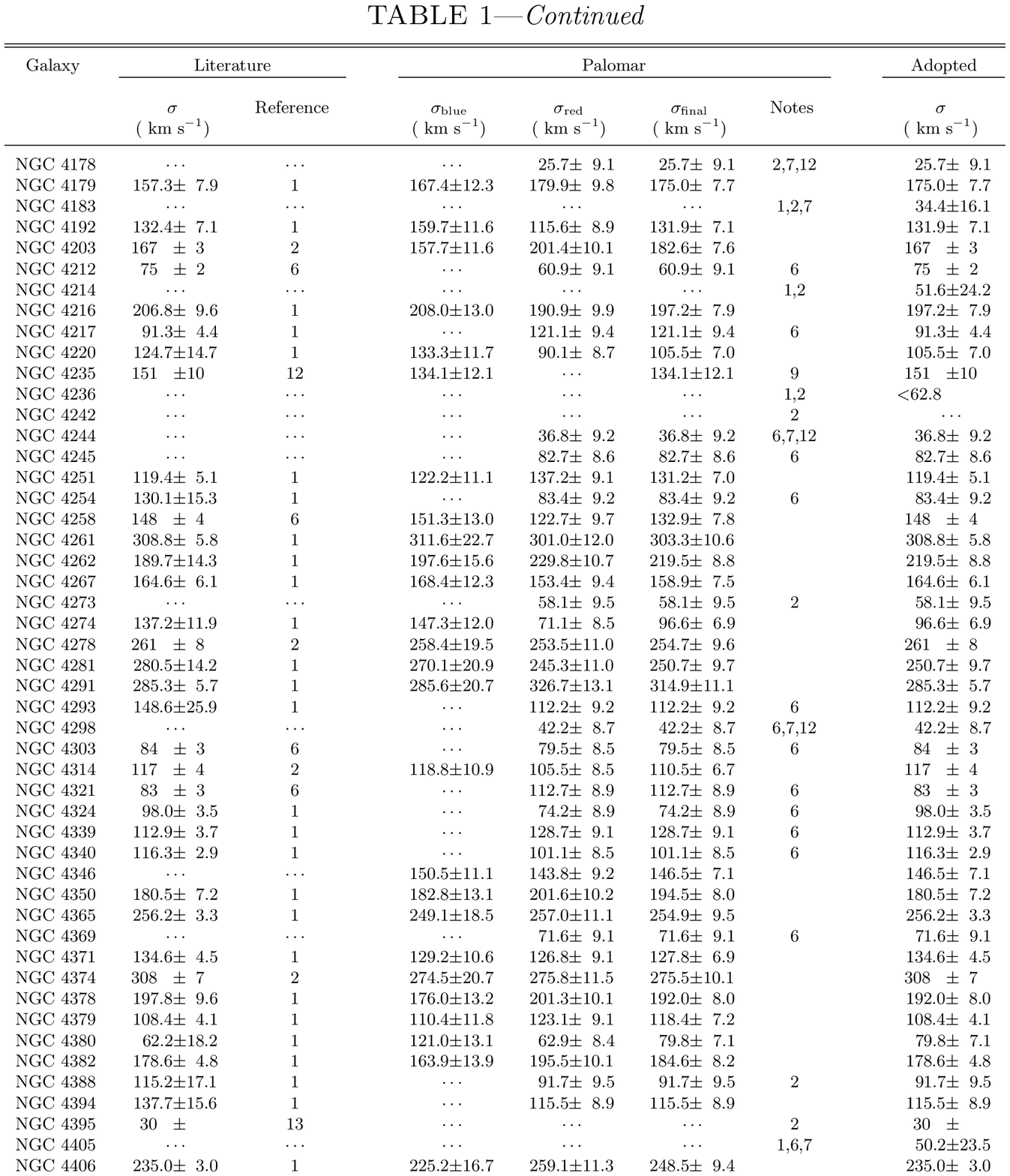,width=18.5cm,angle=0}}
\end{figure*}

\clearpage
\begin{figure*}[t]
\centerline{\psfig{file=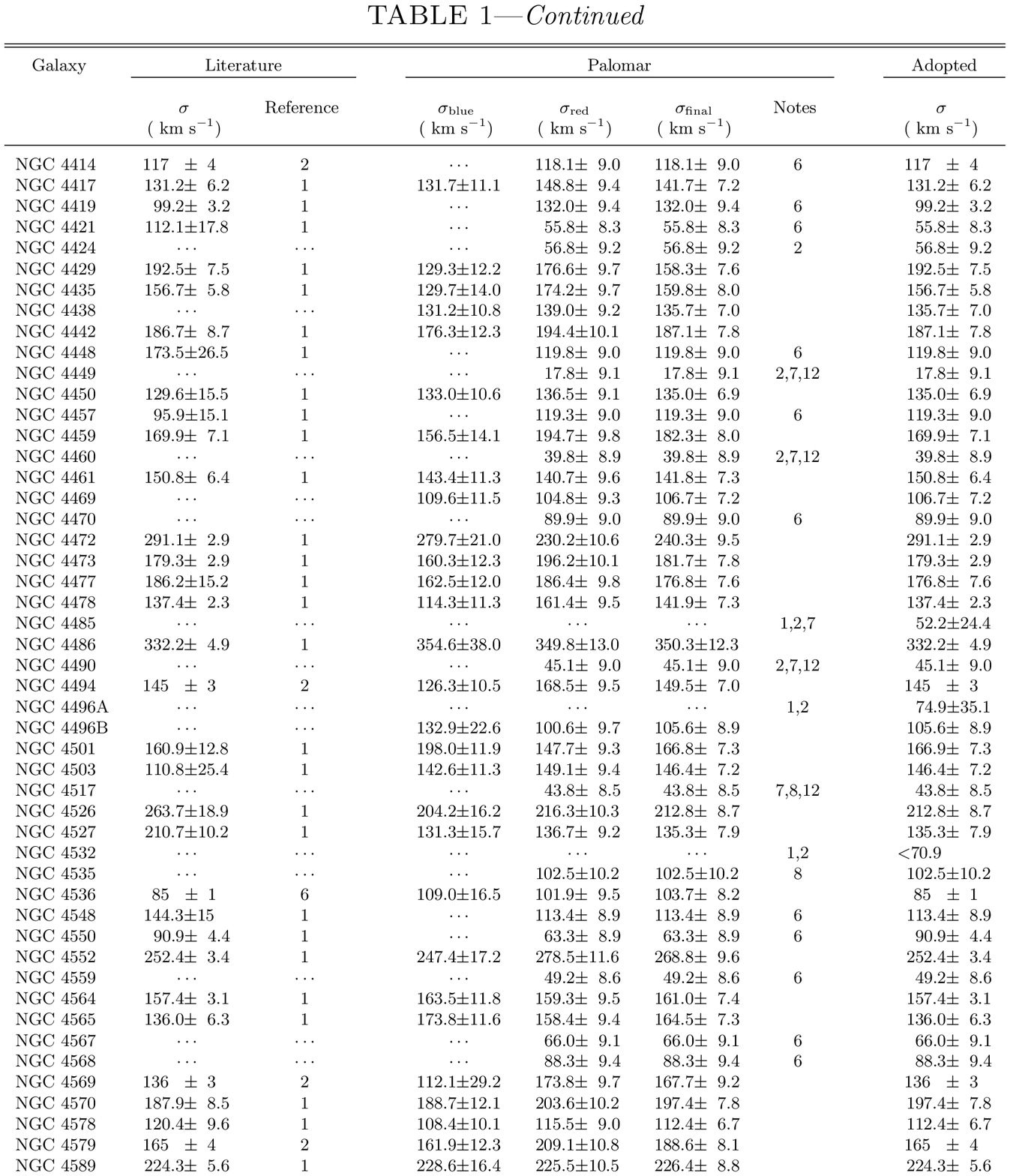,width=18.5cm,angle=0}}
\end{figure*}

\clearpage
\begin{figure*}[t]
\centerline{\psfig{file=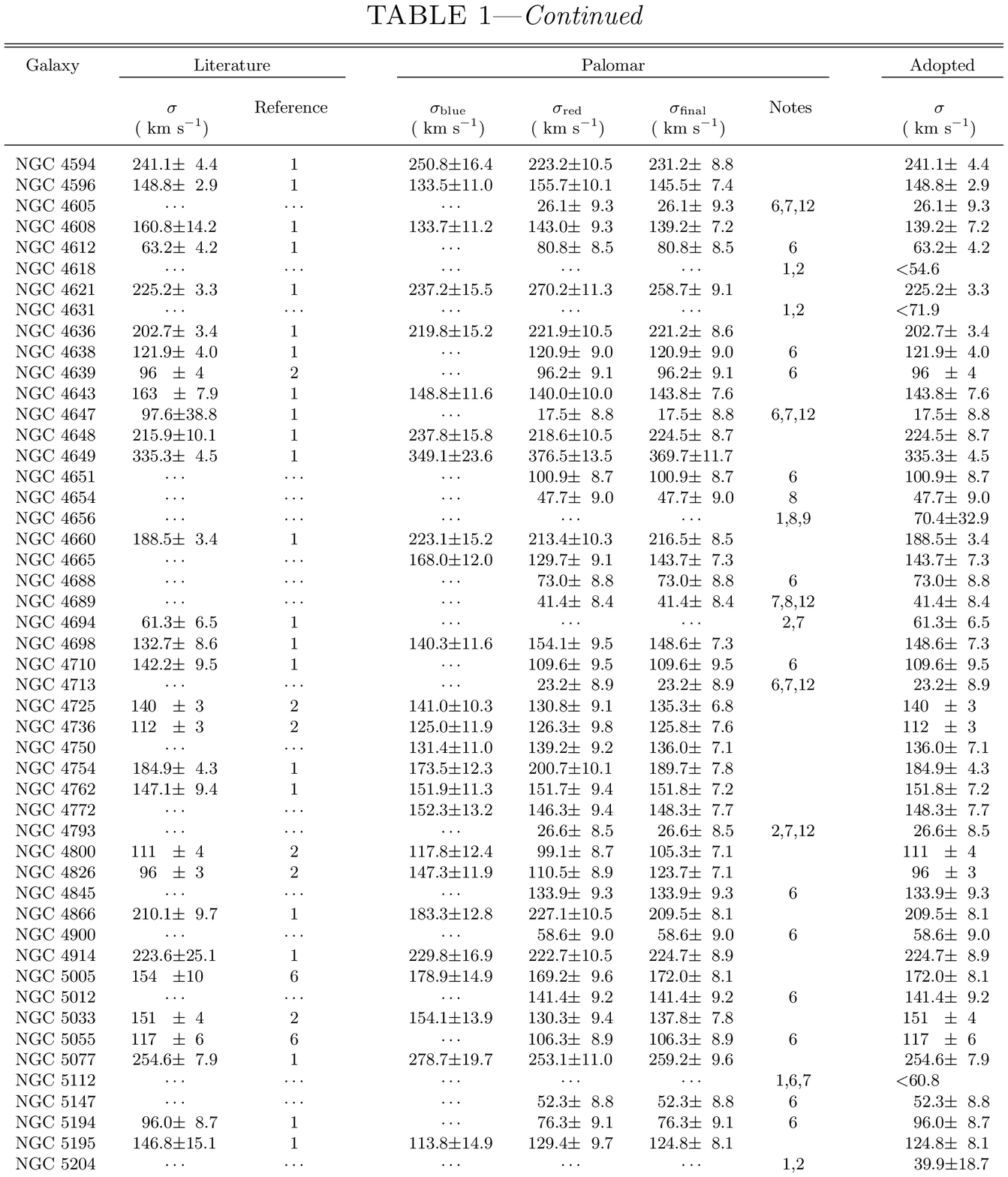,width=18.5cm,angle=0}}
\end{figure*}

\clearpage
\begin{figure*}[t]
\centerline{\psfig{file=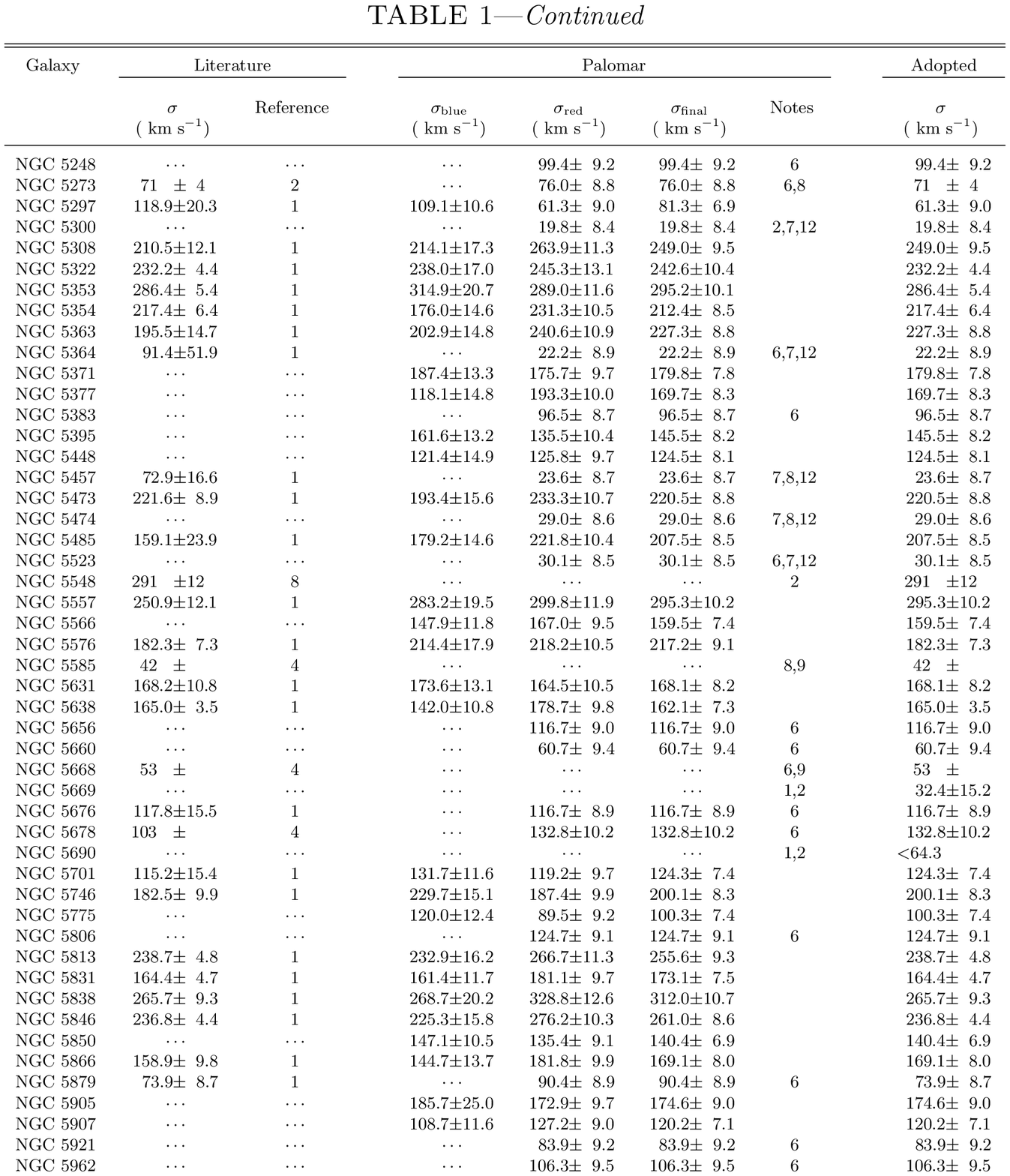,width=18.5cm,angle=0}}
\end{figure*}

\clearpage
\begin{figure*}[t]
\centerline{\psfig{file=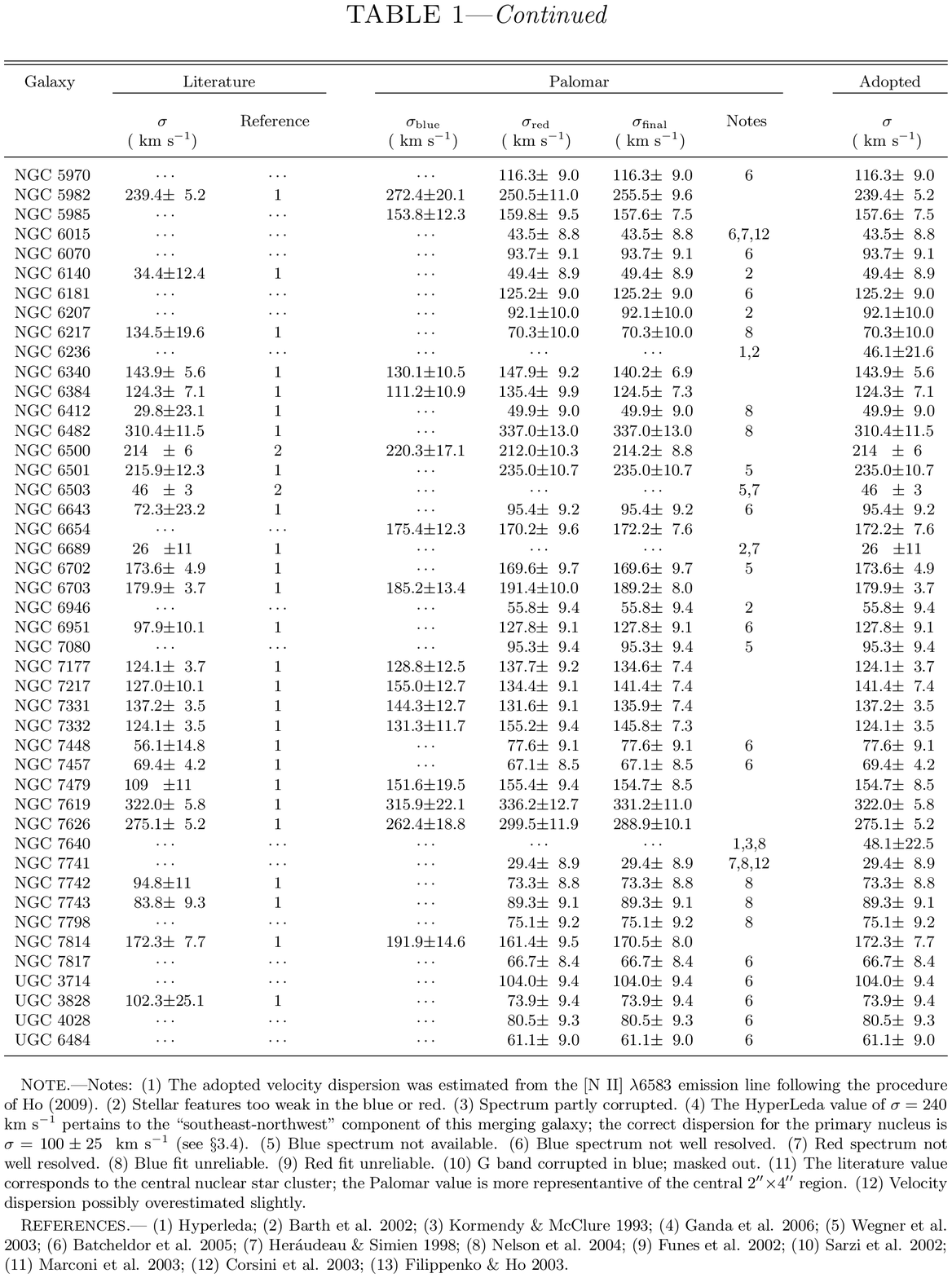,width=18.5cm,angle=0}}
\end{figure*}

\begin{thebibliography}{}

\bibitem[]{} 
Barth, A.~J., Ho, L.~C., \& Sargent, W.~L.~W. 2002, \aj, 124, 2607

\bibitem[]{} 
Batcheldor, D., et al. 2005, \apjs, 160, 76

\bibitem[]{} 
Bender, R. 1990, \aa, 229, 441

\bibitem[]{} 
Bernardi, M., et al. 2003, \aj, 125, 1817

\bibitem[]{} 
Burbidge, E.~M., Burbidge, G.~R., \& Fish, R.~A. 1961, \apj, 133, 393

\bibitem[]{} 
Cappellari, M., \& Emsellem, E. 2004, \pasp, 116, 138

\bibitem[]{} 
Corsini, E.~M., Pizzella, A., Coccato, L., \& Bertola, F. 2003, \aa, 408, 873

\bibitem[]{} 
Davies, R.~L., Burstein, D., Dressler, A., Faber, S.~M., Lynden-Bell, D., 
Terlevich, R.~J., \& Wegner, G. 1987, \apjs, 64, 581

\bibitem[]{} 
de Vaucouleurs, G., de Vaucouleurs, A., Corwin Jr., H.~G., Buta, R.~J., 
Paturel, G., \& Fouqu\'e, R. 1991, Third Reference Catalogue of Bright 
Galaxies (New York: Springer)

\bibitem[]{} 
Filippenko, A.~V., \& Ho, L.~C. 2003, \apj, 588, L13

\bibitem[]{} 
Filippenko, A.~V., Matheson, T, \& Barth, A.~J. 1995, \apj, 455, L139

\bibitem[]{} 
Filippenko, A.~V., Matheson, T., Leonard, D.~C., Barth, A.~J., \&
   Van Dyk, S. D. 1997, PASP, 109, 461

\bibitem[]{} 
Filippenko, A.~V., \& Sargent, W.~L.~W. 1985, \apjs, 57, 503 
 
\bibitem[]{} 
Funes, J.~G., Corsini, E.~M., Cappellari, M., Pizzella, A., Vega Beltr\'an, 
J.~C., \& Bertola, F. 2002, \aa, 388, 50

\bibitem[]{} 
Ganda, K., Falc\'on-Barroso, J., Peletier, R. F., Cappellari, M., Emsellem, 
E., McDermid, R. M., de Zeeuw, P. T., \& Carollo, C. M. 2006, \mnras, 367, 46

\bibitem[]{} 
Graves, G. J., Faber, S. M., Schiavon, R. P., \& Yan, R. 2007, \apj, 671, 243

\bibitem[]{} 
Greene, J. E., \& Ho, L. C.  2006, ApJ, 641, 117

\bibitem[]{} 
H\'eraudeau, Ph., \& Simien, F. 1998, \aas, 133, 317

\bibitem[]{} 
Ho, L.~C. 2008, \annrev, 46, 475

\bibitem[]{} 
------. 2009, \apj, in press

\bibitem[]{} 
Ho, L.~C., Filippenko, A.~V., \& Sargent, W.~L.~W. 1995, \apjs, 98, 477 

\bibitem[]{} 
------. 1997a, \apjs, 112, 315 
 
\bibitem[]{} 
------. 1997b, \apj, 487, 568
 
\bibitem[]{} 
------. 1997c, \apj, 487, 579
 
\bibitem[]{} 
------. 1997d, \apj, 487, 591
 
\bibitem[]{} 
------. 2003, \apj, 583, 159
 
\bibitem[]{} 
Ho, L.~C., Filippenko, A.~V., Sargent, W.~L.~W., \& Peng, C.~Y. 1997e, \apjs,
112, 391

\bibitem[]{} 
Kelson, D.~D., Illingworth, G.~D., van~Dokkum, P.~G., \& Franx, M. 2000, \apj, 
531, 159

\bibitem[]{} 
Kormendy, J., \& McClure, R.~D. 1993, \aj, 105, 1793

\bibitem[]{} 
Marconi, A., et al. 2003, \apj, 586, 868

\bibitem[]{} 
McElroy, D.~B. 1995, \apjs, 100, 105

\bibitem[]{} 
Minkowski, R. 1962, in IAU Symp. 15, Problems of Extra-Galactic Research, ed.
 G.~C. McVittie (New York: Macmillan), 112

\bibitem[]{} 
Morton, D.~C., \& Chevalier, R. 1972, \apj, 174, 489

\bibitem[]{} 
Nelson, C.~H., Green, R. F., Bower, G., Gebhardt, K., \& Weistrop, D. 2004, 
\apj, 615, 652

\bibitem[]{} 
Oke, J.~B., \& Gunn, J.~E. 1982, \pasp, 94, 586

\bibitem[]{} 
Paturel, G., Petit, C., Prugniel, Ph., Theureau, G., Rousseau, J., Brouty, M., 
Dubois, P., \& Cambr\'esy, L. 2003, \aa, 412, 45

\bibitem[]{} 
Prochaska, L. C., Rose, J. A., \& Schiavon, R. P. 2005, AJ, 130, 2666

\bibitem[]{} 
Prugniel, Ph., Zasov, A., Busarello, G., \& Simien, F. 1998, \aas, 127, 117

\bibitem[]{} 
Richstone, D.~O., \& Sargent, W.~L.~W. 1972, \apj, 176, 91

\bibitem[]{} 
Rix, H.-W., \& White, S.~D.~M. 1992, \mnras, 254, 389

\bibitem[]{} 
Sandage, A.~R., \& Tammann, G.~A. 1981, A Revised Shapley-Ames Catalog of
Bright Galaxies (Washington, DC: Carnegie Inst. of Washington)

\bibitem[]{} 
Sargent, W.~L.~W., Schechter, P.~L., Boksenberg, A., \& Shortridge, K. 1977,
\apj, 212, 326

\bibitem[]{} 
Sarzi, M., et al. 2002, \apj, 567, 237

\bibitem[]{} 
Simkin, S.~M. 1974, \aa, 31, 129

\bibitem[]{} 
Stanford, S.~A., \& Balcells, M. 1990a, in IAU Colloq. 124, Paired and 
Interacting Galaxies, ed. J. W. Sulentic, W. C. Keel \& C. M. Telesco
(Washington, D. C.: NASA Conference Publications No. 3098), 347

\bibitem[]{} 
------. 1990b, \apj, 355, 59

\bibitem[]{} 
Statler, T.~S. 1995, \aj, 109, 1371

\bibitem[]{} 
Tonry, J., \& Davis, M. 1979, \aj, 84, 1511

\bibitem[]{} 
Valdes, F., Gupta, R., Rose, J. A., Singh, H. P., \& Bell, D. J. 2004, \apjs, 
152, 251

\bibitem[]{} 
van~der~Marel, R.~P. 1994, \apj, 270, 271

\bibitem[]{} 
van der Marel, R.~P., \& Franx, M. 1993, \apj, 407, 525

\bibitem[]{} 
Wegner, G., et al. 2003, \aj, 126, 2268

\bibitem[]{} 
Whitmore, B.~C., McElroy, D.~B., \& Tonry, J.~L. 1985, \apjs, 59, 1

\bibitem[]{} 
Zhang, Y., Gu, Q.-S., \& Ho, L. C. 2008, \aa, 487, 177

\end{thebibliography}
\end{document}